\documentclass[preprint]{aastex}

\shorttitle{ON  $^7$Li ENRICHMENT BY LOW MASS METAL POOR}
\shortauthors{de la Reza et al.}

\begin{document}

\title  {ON  $^7$Li ENRICHMENT BY LOW MASS METAL POOR 
                                             RED GIANT BRANCH STARS}

\author{Ramiro de la Reza\altaffilmark {1}, Licio da Silva\altaffilmark {1},
        Natalia  A. Drake\altaffilmark {1,2} \& Marco A. Terra\altaffilmark {3}}

\affil{$^{1}$Observat\'orio Nacional - Rio de Janeiro - Brazil}
\affil{$^{2}$Astronomical Institute, St. Petersburg State University, St. Petersburg, 
              Russia}
\affil{$^{3}$Observat\'orio do Valongo -  UFRJ - Rio de Janeiro - Brazil}
\email{delareza@on.br}

\begin{abstract}
First-ascent red giants with strong and very strong Li lines have just 
been discovered in
globular clusters. Using the stellar internal prompt $^7$Li enrichment -
mass loss scenario, we explore the possibility of $^7$Li enrichment in the interstellar
matter of the globular cluster M3 produced by these Li rich giants.
We found that enrichment as large as 70\% or more
compared to the initial $^7$Li content of M3 can be obtained  during the entire life 
of this cluster. However, due to the several crossings of 
M3 into the Galactic plane, the new $^7$Li will be very probably removed by ram pressure
into the disk. 
Globular clusters appear then as possible new sources of $^7$Li in
the galactic disk. It is also suggested that the known Na/Al variations 
in stars of globular clusters
could be somehow related to the $^7$Li variations and that the cool bottom 
mixing
mechanism acting in the case of $^7$Li could also play a role in the case
of Na and Al surface enrichments.

\end{abstract}


\section {INTRODUCTION}

The main purpose of this letter is to discuss the possibility of 
$^7$Li enrichment in some components of the Galactic halo in view 
of recent discoveries of very strong and strong Li giant 
stars in globular clusters M3 \citep{kra99}, 
NGC$\,$362  \citep{smi99}
and M5 \citep{car98}
In general, Galactic chemical 
evolution models contain the following sources of  $^6$Li and $^7$Li 
enrichment in the halo: Galactic cosmic rays $\alpha - \alpha$  nucleosynthesis, SN$\,$II  
$\nu$-process and the $^7$Be process in intermediate mass ($3M_\odot < M < 7M_\odot$)  AGB stars  
\citep{del95,boe98,dan91,rom99}.
The $^6$Li
detection in two stars HD$\,$84937 \citep{smi93,smi98}
and BD$+26^\circ 3578$ \citep{hob94,hob97} 
indicated for the first time that a Galactic source 
other than a cosmologic one was at play, at least in the metallic range of the
interstellar matter preceding the 
formation of stars having [Fe/H]\footnote{here we use the following notation   
[X/H]$ = \log ({\rm X/H})_* - \log ({\rm X/H})_\odot$ and 
$\log\epsilon_{\rm X }= \log({\rm X/H}) + 12$, where X and H are the number of 
atoms of the element X and hydrogen respectively} $ \sim - 2.3$. Cosmic rays have  
produced $^6$Li in a pre-stellar interstellar medium by $\alpha - \alpha$ process or 
directly on stars. This last possibility has been shown, however, to be inefficient 
\citep{lam95}.
Also, the creation of Li on stars' surfaces appears to be uncertain  
\citep{mon98}.
Difficulties related to $^7$Li formation by SN$\,$II $\nu$  and AGB giants are 
discussed by \citep{boe98}.
They concern the absence of spectral signatures 
in the Li enriched dwarf and sub-giant stars proving the action of these processes such as the 
absence of excesses of Mg in the case of  SN$\,$II and of the $s$-elements 
in the case of AGB stars. 
These Li-rich objects as the field star BD$+23^\circ 3912$ \citep{king96}
present an overabundance in respect to the mean $^7$Li 
abundance of the Spite plateau \citep{spi82}
Observational $^7$Li enrichment 
in AGB giants which have been discovered in the LMC \citep{smi89,smi90}
are, however, in a much sounder theoretical basis \citep{sac92,maz99}.
It appears interesting to explore 
the possibilities of a new source of  $^7$Li enrichment in population II stars systems. 
This source is the metal poor low mass $(M < 2.5 M_\odot)$ 
first-ascent red giant branch stars (RGB).

\section{THE METAL DEFICIENT GIANTS $^7$Li ENRICHMENT PROCESS}

Low metal RGB stars are expected to lose a reasonable quantity of mass 
$(0.1 - 0.2M_\odot)$ during their first ascent on the giant branch. 
Considering a time scale to reach the tip of this branch as 
$2\times 10^6$ y \citep{roo72,kra93}
we obtain mean stellar mass 
losses of the order of $5\times10^{-8}$ up to $10^{-7} M_\odot$/y. 
It is interesting to note that the same mass loss rates 
are present, nevertheless in a discontinued way in the mass loss -- $^7$Li enrichment scenario 
proposed by \citet{rez96}
and \citet{rez97}.
In this 
scenario, all low mass giants $(M < 2.5M_\odot$) suffer a prompt $^7$Li 
enrichment in the upper 
part of the RGB, after the first dredge-up and before the RGB tip. 
The internal 
mechanism producing  new $^7$Li forms a circumstellar shell (CS) enriched with 
$^7$Li, which detaches from the star when the internal process ceases. 
In this way, the interstellar medium is enriched with $^7$Li. 
The synchronized expansion of the dusty CS 
enriched with $^7$Li and the subsequent $^7$Li depletion in star photospheres can be 
followed by means of closed loops in an IRAS color-color diagram 
and compared with observed positions of the stars \citep{rez96,rez97}.
These times of expansion measure the ``lithium cycles'' which are the periods
when strong Li lines are observed. Li cycles of the order of 
$10^3$ up to $10^5$ y have been estimated for an
CS expanding velocity of 2 km/s \citep{rez97}.
Maybe the best mechanism for producing the  $^7$Li photospheric 
enrichment from internal origin for these low mass 
giants is the ``cool bottom process'' (CBP) \citep{sac99}
based on the $^7$Be production by means of the excess of $^3$He 
in the H-burning shell which is characteristic of these low mass stars. 
The fresh $^7$Be is transported by a conveyor circulating mechanism up to 
the base of the convective layer to be then taken to the stellar 
surface where it is transformed into $^7$Li. 
Because the mentioned conveyor mechanism attains 
deeper and hotter regions in the case of metal deficient giants, very 
large $^7$Li surface abundances  ($\log\epsilon_{\rm Li} \sim
4.2$) have been obtained for [Fe/H]$ = - 2.3$ in a  short episode (see 
Fig.~10 in \citet{sac99}.
In the above mentioned scenario which links a discontinuous $^7$Li enrichment 
to mass loss, the process can  be 
repeated depending on the available quantity of $^3$He. The nature of the 
internal process which provokes the formation of the CS is not specified however; some 
mechanisms can be considered, such  as a conversion of a rapidly rotating 
core \citep{fek93,fek96}
or an outward extension of large 
convection cells as those which appear to be producing the extended atmosphere of Betelgeuse
\citep{lim98}.

Another way to examine the Li enrichment is by relating it
to the luminosity bump in the RGB stage. This bump corresponds to the evolutionary
stage when the hydrogen burning shell erases the chemical discontinuity left behind by
the first dredge-up at the moment the convective envelope was at its maximum extent 
or deepest penetration. Because the CBP is assumed to start at this stage until the tip
of the RGB is reached, we expect, if the CBP is responsible for the $^7$Li enrichment, to
observe Li-rich giants at luminosities equal or higher than the RGB bump. 
This is the case for the giant IV-101 in M3 where the star is observed at 
$V = 13.2^{\rm m}$ and the very precise
RGB bump of M3 is $V = 15.45^{\rm m} \pm 0.05^{\rm m}$ (Ferraro et al. 1999).
Concerning population I Li K giants, and especially
those stars having Hipparcos parallaxes, \citet{rez00}
has shown that these giants have higher luminosities than those corresponding 
to the bumps proposed by \citet{cha94}
at least for stars with masses between 1.0 to $1.5M_\odot$.

The relation of the $^7$Li and $^{13}$C enrichment is less clear, 
principally because of the scarce
number of Li rich stars among the metal poor giants. 
In general, among population I Li K giants \citet{sil95}
and  \citet{dra98}
have not found a clear relation between the Li
abundances and the $^{12}$C/$^{13}$C ratios indicating that 
probably both enrichments are not time correlated. 
As far as the RGB bump is concerned, \citet{cha98}
have shown 
that the $^{13}$C enrichment is found only at the bump or 
higher luminosities when considering the 
$^{12}$C/$^{13}$C ratios measured by 
Shetrone et al. (1993) 
in stars with $-1.0 \le {\rm [Fe/H]} \le -0.5$.
Depending on the metallicity, population II giants can be sources of $^7$Li only in the RGB stage
and not in the following AGB stage \citep{sac99}.
Due to their higher hydrogen burning temperatures, very metal 
poor giants at [Fe/H] $\sim -2.3$ for example, not only produce
higher quantities of fresh $^7$Li, but also destroy all the $^3$He 
during the RGB stage so that no more of it remains to produce $^7$Li in the AGB stage. 
As far as  the very Li-rich giant star IV-101 in M3 is concerned, 
it  does not show any particular enhancement 
of the $s$-process elements, presenting then a typical RGB scenario. 
Due to incomplete $^3$He burning, low mass  mild deficient stars will maintain 
$^7$Li creation by CBP during the AGB stage.

Concerning population I giants, the distinction between RGB and 
AGB contributions of $^7$Li will 
depend more on the stellar mass. Low mass stars, below $2.5M_\odot$, 
produce $^7$Li  in the RGB and the AGB by CBP \citep{sac99},
 whereas giants with masses between 4 and $7M_\odot$ contribute in the 
AGB by means of the Hot Bottom Process \citep{sac92,maz99}.
The CBP needs to be explored to see if it can explain 
the intermediate mass zone between 2.5 and $4M_\odot$. 

To evaluate the interstellar Li enrichment in the prompt $^7$Li 
scenario presented before, it is fundamental
to know the maximum Li abundances for stars presenting fresh 
$^7$Li before depletion begins to act. We consider that
this is the case for IV-101 and for estimating the $^7$Li enrichment 
in M3 as will be presented in a next section, we have
calculated the Li abundance of IV-101 in Non-LTE (NLTE) by means of the 
$\lambda6708$ and $\lambda6104$ \ion{Li}{1} lines using a new self-consisting
methodology for chromospheric treating. Details of this method can be found in 
\citet{ter97}
and will be submitted elsewhere \citep{ter00}.
The obtained $^7$Li abundance for IV-101 is log $\epsilon_{\rm Li}$ = 4.0.
This NLTE value is an order of magnitude larger than the LTE Li 
abundance proposed by \citet{kra99}
based on the $\lambda6708$ resonant Li line alone.

\section  {IS THERE A $^7$Li - Na/Al  ENRICHMENT CONNECTION?}

An extensive literature exists on the CNO, Na and Al variations among globular cluster stars  
(a review can be seen, for instance, in \citet{kra94}). The most 
remarkable are the Na and Al versus O anticorrelation and the Na and Al versus N 
correlation. These variations indicate that a relatively rapid mixing is taking place. 
Considering this, we can ask if there is a $^7$Li -- Na  and a $^7$Li -- Al simultaneous 
surface enrichment. \citet{kra99}
have suggested that a Li -- Al
correlation could be present in some giants of metallicity ${\rm [Fe/H]} = -1.5$ in M3. 
We did not find a similar correlation for mild deficient giants 
having extremely  high $^7$Li abundances as is the case of  the 
high velocity star PDS$\,$68 ([Fe/H]$ = -0.4$) 
where no substantial Na enrichment was found \citep{dra98}.
Substantial Li -- Al/Na correlations must exist in very low metal RGB stars because at 
[Fe/H]  $\sim - 2.3$ relatively  similar internal star regions  produce large increases of  
$^{23}$Na and $^{27}$Al  (from the seed elements $^{20}$Ne and 
$^{24}$Mg respectively) 
(Cavallo, Sweigart, \& Bell 1996) 
and that of $^7$Li (from the seed element $^3$He) \citep{sac99}.
The 
cool bottom processing proposed by Sackmann \& Boothroyd for low mass giants producing the large surface 
Li enrichments could also play an  important role concerning the Na and Al enrichment variations.
If this is the case, a $^7$Li -- Al/Na connection will be independent from any stellar interactions
due to high stellar density  in globular clusters and will 
be also valid  for field stars. \citet{fuj99}
have suggested another scenario
to explain Na/Al variations (not $^7$Li) in globular cluster stars  by means of shell flashes
induced by a deep H mixing provoked by star -- star interactions in a dense cluster.

\section  {ON THE $^7$Li PRODUCTION IN M3}

We calculate here the enrichment of new $^7$Li in the interstellar matter of 
M3 produced by low mass giant stars $(M < 2.5M_\odot$). 
These stars are considered being a second
generation stars formed by matter already enriched in a large
part of heavy elements by a first generation of high mass, short life, stars.
The prompt $^7$Li enrichment - mass loss scenario used here will get into
action only when the first giants of mass $\sim 2.5M_\odot$ appear in M3. That
is after $10^8$ y, which is a small fraction of the lifetime of the globular cluster. 
The $^7$Li production by the $\sim2.0M_\odot$ stars will continue during an important
part of 
the life of M3. Later, the $^7$Li production will increase in time due 
to the rise of population of low mass giants ($\sim 1M_\odot$). 
Due to time evolution constraints, the $^7$Li prompt enrichment -- mass loss
mechanism will very probably never be able to operate during any initial
self-enrichment of the gas cloud from which the present M3 cluster was formed.
Some recent results \citep{dra98}
indicate that $^7$Li photospheric depletion, following a strong 
enrichment, depends on the value of the stellar mass for masses less 
than $2.5M_\odot$. RGB stars with masses $\sim  2M_\odot$ have larger 
depletion times ($\ge 10^4$ y) than those 
of stellar masses of $\sim 1M_\odot$ ($\sim 3 \times 10^3$ y). These results were obtained 
for mass losses  between $10^{-8} - 10^{-7} M_\odot/$y and CS expansion velocities 
equal to 2 km/s and for [Fe/H] between $-0.5$ and $0.2$. 
In the case of population II giants $(M < 2.5M_\odot$), 
where stellar masses around  $1M_\odot$ are of interest, we obtain short Li cycles 
resulting in lower probabilities of detection. Considering the time to reach the RGB tip 
as $2\times10^6$ y, the probability to detect a Li-rich RGB star will be $\sim 3 \times 10^3\, {\rm y}/ 
2 \times 10^6\, {\rm y}$, that is 0.15\%. However, if the $^7$Li enrichment process is repeated by a 
recurrence factor of 10 for example, we obtain 1.5\%. A similar result is found by Kraft 
et al. (1999)  considering that already two Li-rich RGB giants have been found among 
near 100 observed globular clusters giants. We must note that the physical basis for the actual value 
of the recurrence factor depends on the quantity of $^3$He that remains to be burned.
Let us estimate the $^7$Li enrichment in the interstellar medium in the 
globular cluster M3 by means of mass loss of  super Li-rich RGB stars, following the scenario of 
\citet{rez96,rez97}.
Considering that all RGB stars are potential sources of $^7$Li 
during the short time in which they form and eject a  Li-rich CS, 
the total production of $^7$Li will be 
$P_{\rm Li} = N_{\rm RGB} \cdot N_{\rm Li} \cdot f \cdot t_{\rm CS} \cdot \dot M / t_{\rm RGB}$. 
Here  $N_{\rm RGB}$ is 
the total number of RGB stars in the globular cluster M3, 
$N_{\rm Li} = (n_{\rm Li}/n_{\rm H})\cdot (m_{\rm Li}/m_{\rm H})$ where $(n_{\rm Li}/n_{\rm H})$ is the ratio 
of the number of Li and H atoms equal to 
$10^{-8}$ for $\log \epsilon_{\rm Li} = 4.0$ and $m_{\rm Li} = 7 m_{\rm H}$,  $f$ is the 
recurrence factor, $t_{\rm CS} = 200$ y is the time of CS  formation  
(de la Reza et al. 1996), 
$\dot M$ is the  mass loss equal to  $10^{-7}M_\odot/$y and $t_{\rm RGB}$ is the time 
necessary to attain the RGB tip ($2 \times 10^6$ y). The number of RGB stars ($N_{\rm RGB}$) can be estimated in 
the following way: 
we assume that the present observed $N_{\rm RGB}$ will represent a 
reasonable mean value of the number of RGB stars during the entire life of the globular 
cluster. 
(We consider that RGB stars have typical 
masses around the turn-off mass and similar to the one assumed  
by \citet{kra99}
for IV-101 ($0.85M_\odot$)). 
Probably a better evaluation of the production of $^7$Li during the 
life of M3 can be made, if we are able to distinguish the Li yields 
produced by giants with masses around $\sim 2.0M_\odot$ from those with 
  $\sim 1.0M_\odot$, which is not unfortunately the case. Even in a crude way, 
$N_{\rm RGB}$ can be estimated by extrapolating the already 
known number of RGB stars (424) counted among almost 19$\,$000 stars in M3 \citep{fer97}.
Maintaining the same proportion for the total estimated number of stars in M3 equal to
 $3.44 \times 10^5$ stars (Lang 1992) we obtain approximately 7700 RGB 
stars in M3. Taking first $f = 1$ we obtain $P_{\rm Li} = 
5.4 \times 10^{-15}M_\odot$/y of new $^7$Li in the gas of the globular cluster interstellar 
medium. If we multiply this value by the mean age of the globular clusters (13 Gyrs, \citet{mou98})
and if we consider this age to be that of M3, we obtain a total 
$^7$Li mass production of $7.0 \times 10^{-5}$$M_\odot$. To have a better 
idea of what represents this quantity of new $^7$Li let us compare the latter with the initial content 
of $^7$Li in M3. To estimate this initial quantity of $^7$Li 
we can consider a Jeans mass of $10^6 M_\odot$ at the 
earliest time of formation for M3 \citep{pee93}.
Supposing for the sake of simplicity, that almost all 
matter consists of H atoms and considering an initial $^7$Li 
abundance of $n_{\rm Li}/n_{\rm H} = 1.6 \times 10^{-10}$ corresponding to the 
Spite plateau \citep{mol95,bon97},
we obtain an initial mass of $^7$Li of $ 10^{-3} M_\odot$ in M3. The RGB 
production will represent only the  $\sim$ 7\% of this
quantity. But if we consider $f = 10$ this value will increase up to $\sim 70$\%. 
If we consider alternatively possible higher  $^7$Li abundances as 
$\log \epsilon_{\rm Li} = 4.5$, as obtained for some giants such as PDS$\,$68, and 
if we use a larger, but yet realistic, mass loss value  of
$5\times 10^{-7} M_\odot/$y,  we will obtain
increased enrichment factors.       
It is interesting to note, that  this new $^7$Li in the interstellar 
gas of M3 will, very probably, not be maintained for long time 
in the cluster and will be transferred, by ram pressure, to the 
Galactic disk when M3 crosses this one.
\citet{Scholz93}
have calculated that M3 crossed the plane nearly 34 times during 
the last 10 Gyrs. Globular clusters appear 
then as new potential sources of $^7$Li enrichment in the  Galactic disk!

\acknowledgments
We thanks the referee for very useful suggestions which clarified some
aspects of this letter. N.A.D. and M.A.T. thank FAPERJ for the financial support under the 
contracts E-26/151.172/98 and E-26/150.571/98 respectively. RdlR and LdS thank CNPq by the
financial support grants 301375/86-0 and 200580/97-0 respectively.

\end{document}